\documentclass[twoside,twocolumn,9pt]{article}
\usepackage{extsizes}
\usepackage[super,sort&compress,comma]{natbib} 
\usepackage[version=3]{mhchem}
\usepackage[left=1.5cm, right=1.5cm, top=1.785cm, bottom=2.0cm]{geometry}
\usepackage{balance}
\usepackage{times,mathptmx}
\usepackage{sectsty}
\usepackage{graphicx} 
\usepackage{lastpage}
\usepackage[format=plain,justification=justified,singlelinecheck=false,font={stretch=1.125,small,sf},labelfont=bf,labelsep=space]{caption}
\usepackage{float}
\usepackage{fancyhdr}
\usepackage{fnpos}
\usepackage[english]{babel}
\addto{\captionsenglish}{%
  
}
\usepackage{array}
\usepackage{droidsans}
\usepackage{charter}
\usepackage[T1]{fontenc}
\usepackage[usenames,dvipsnames]{xcolor}
\usepackage{setspace}
\usepackage[compact]{titlesec}
\usepackage{hyperref}

\definecolor{cream}{RGB}{222,217,201}

\begin{document}

\pagestyle{fancy}
\thispagestyle{plain}
\fancypagestyle{plain}{

\renewcommand{\headrulewidth}{0pt}
}

\makeFNbottom
\makeatletter
\renewcommand\LARGE{\@setfontsize\LARGE{15pt}{17}}
\renewcommand\Large{\@setfontsize\Large{12pt}{14}}
\renewcommand\large{\@setfontsize\large{10pt}{12}}
\renewcommand\footnotesize{\@setfontsize\footnotesize{7pt}{10}}
\makeatother

\renewcommand{\thefootnote}{\fnsymbol{footnote}}
\renewcommand\footnoterule{\vspace*{1pt}%
\color{cream}\hrule width 3.5in height 0.4pt \color{black}\vspace*{5pt}} 
\setcounter{secnumdepth}{5}

\makeatletter 
\renewcommand\@biblabel[1]{#1}            
\renewcommand\@makefntext[1]%
{\noindent\makebox[0pt][r]{\@thefnmark\,}#1}
\makeatother 
\renewcommand{\figurename}{\small{Fig.}~}
\sectionfont{\sffamily\Large}
\subsectionfont{\normalsize}
\subsubsectionfont{\bf}
\setstretch{1.125} 
\setlength{\skip\footins}{0.8cm}
\setlength{\footnotesep}{0.25cm}
\setlength{\jot}{10pt}
\titlespacing*{\section}{0pt}{4pt}{4pt}
\titlespacing*{\subsection}{0pt}{15pt}{1pt}

\fancyfoot{}
\fancyfoot[LO,RE]{\vspace{-7.1pt}\includegraphics[height=9pt]{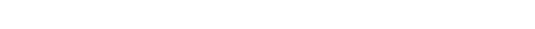}}
\fancyfoot[CO]{\vspace{-7.1pt}\hspace{11.9cm}\includegraphics{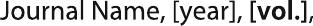}}
\fancyfoot[CE]{\vspace{-7.2pt}\hspace{-13.2cm}\includegraphics{head_foot/RF}}
\fancyfoot[RO]{\footnotesize{\sffamily{1--\pageref{LastPage} ~\textbar  \hspace{2pt}\thepage}}}
\fancyfoot[LE]{\footnotesize{\sffamily{\thepage~\textbar\hspace{4.65cm} 1--\pageref{LastPage}}}}
\fancyhead{}
\renewcommand{\headrulewidth}{0pt} 
\renewcommand{\footrulewidth}{0pt}
\setlength{\arrayrulewidth}{1pt}
\setlength{\columnsep}{6.5mm}
\setlength\bibsep{1pt}

\makeatletter 
\newlength{\figrulesep} 
\setlength{\figrulesep}{0.5\textfloatsep} 

\newcommand{\topfigrule}{\vspace*{-1pt}%
\noindent{\color{cream}\rule[-\figrulesep]{\columnwidth}{1.5pt}} }

\newcommand{\botfigrule}{\vspace*{-2pt}%
\noindent{\color{cream}\rule[\figrulesep]{\columnwidth}{1.5pt}} }

\newcommand{\dblfigrule}{\vspace*{-1pt}%
\noindent{\color{cream}\rule[-\figrulesep]{\textwidth}{1.5pt}} }

\makeatother

\twocolumn[
  \begin{@twocolumnfalse}
{\includegraphics[height=30pt]{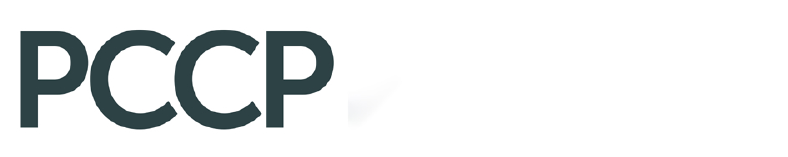}\hfill\raisebox{0pt}[0pt][0pt]{\includegraphics[height=55pt]{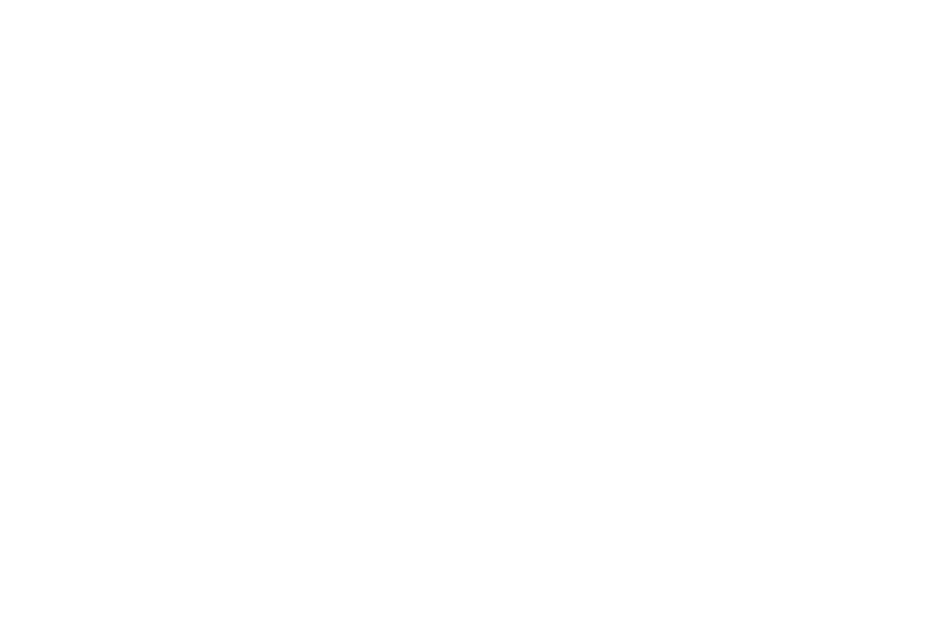}}\\[1ex]
\includegraphics[width=18.5cm]{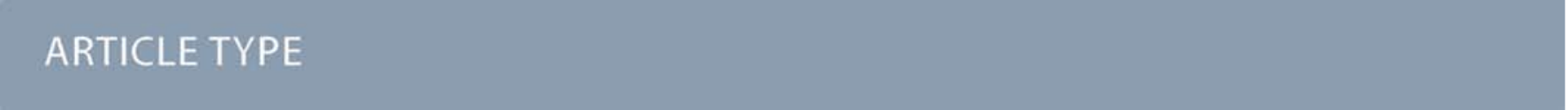}}\par
\vspace{1em}
\sffamily
\begin{tabular}{m{4.5cm} p{13.5cm} }

\includegraphics{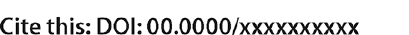} & \noindent\LARGE{\textbf{Supramolecular Chemistry Based on 4-Acetylbiphenyl on Au(111)$^\dag$}} \\
\vspace{0.3cm} & \vspace{0.3cm} \\

 & \noindent\large{Roberto Robles,$^{\ast}$\textit{$^{a}$} Vladim\'{i}r Zoba\v{c},\textit{$^{a}$} Kwan Ho Au Yeung,\textit{$^{b}$} Francesca Moresco,\textit{$^{b}$} Christian Joachim,\textit{$^{c}$} and Nicol\'as Lorente\textit{$^{a,d}$}} \\

\includegraphics{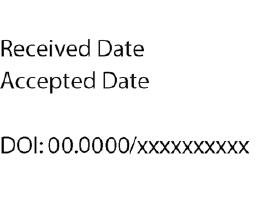} & \noindent\normalsize{On a gold surface, supramolecules composed of 4-acetylbiphenyl molecules show structural directionality, reproducibility and robustness to external perturbations. We investigate the assembly of those molecules on the Au(111) surface and analyze how the observed supramolecular structures are the result of weak long-range dispersive forces stabilizing the 4-acetylbiphenyl molecules together. Metallic adatoms serve as stabilizing agents. Our analysis suggests new ways of creating complex molecular nano-objects that can eventually be used as devices or as seeds for extended hierarchical structures.} \\

\end{tabular}

 \end{@twocolumnfalse} \vspace{0.6cm}

  ]

\renewcommand*\rmdefault{bch}\normalfont\upshape
\rmfamily
\section*{}
\vspace{-1cm}


\footnotetext{\textit{$^{a}$~Centro de F\'{i}sica de Materiales CFM/MPC (CSIC-UPV/EHU), Paseo de Manuel de Lardizabal 5, 20018 Donostia-San Sebasti\'an, Spain. E-mail: roberto.robles@ehu.es}}
\footnotetext{\textit{$^{b}$~Center for Advancing Electronics Dresden, TU Dresden, 01069 Dresden, Germany }}
\footnotetext{\textit{$^{c}$~Centre d'Elaboration de Mat\'eriaux et d'\'Etudes Structurales (CEMES), CNRS, Universit\'e de Toulouse, 29 Rue J. Marvig, BP 4347, 31055 Toulouse Cedex, France}}
\footnotetext{\textit{$^{d}$~Donostia International Physics Center (DIPC), 20018 Donostia-San Sebasti\'an, Spain}}

\footnotetext{\dag~Electronic Supplementary Information (ESI) available: Experimental image of the surface covered by supramolecules; description of the interaction of Au adatoms with tetramers; Au adatoms away from the center of the supramolecule; simulation results for ABP trimers; simulation results of NEB energy barrier calculations. See DOI: 10.1039/cXCP00000x/}


\section{Introduction}
Supramolecular chemistry describes the synthesis of molecular structures where intermolecular binding forces between the different parts are smaller than the intramolecular ones. The non-covalent bonds can go from hydrogen to halogen bonds, combined with weak electrostatic interactions. In that respect, the interaction with the supporting surface can give rise to unique supramolecular assemblies. The surface plays here two mandatory roles: (i) making molecular weak reactants to meet because the diffusion process is constrained to two dimensions (2D) and (ii) mediating molecular interactions.\cite{amabilino_supramolecular_2016,clair_controlling_2019} It results in a rich variety of hierarchical molecular self-organization on surfaces that is the object of many efforts.\cite{lorenzo_extended_2000,kuhnle_chiral_2002,barth_engineering_2005,barth_molecular_2007,cirera_thermal_2016} There is an intermediate soft chemistry regime between this on-surface 2D self-arrangement and the single isolated molecule limit. Here, supramolecular structures with a finite number of molecules can assemble giving rise to a new class of molecular objects on a surface. Examples are weakly-bound molecular clusters. They have the same properties in terms of mechanical stability as large covalently bond molecules, but they are only found on a surface and they are easily separated into their initial molecular chemical groups. This is the case of functionalized biphenyl molecules.\cite{kong_long-range_2020, nickel_moving_2013,ohmann_supramolecular_2015} Using hydroxyl (OH) lateral groups, strong hydrogen bonds take place between the partner molecules on the supporting surface. Uniquely, a single type of chiral molecular assemblage composed of 3 of those molecules forms by optimizing the H-bonding orientation. When the coverage increases, a long-range order sets in forming self-organized molecular coverage of the surface.\cite{kong_long-range_2020} If acetyl (COCH3) groups are used instead of OH, a rich variety of molecular assemblage can be formed, going from two to four molecules in the supramolecule.\cite{nickel_moving_2013} Once formed, the supramolecules are all identical and can be studied and manipulated by scanning tunneling microscope (STM).\cite{nickel_moving_2013,ohmann_supramolecular_2015,eisenhut_training_2016} 

Previous studies\cite{meyer_tuning_2015} have also shown the importance of single gold adatoms for the formation of on-surface supramolecules. It was shown that a gold atom could interact with the cyano group of three cyanosexiphenyl molecules that are thus on-surface stabilized by this adatom. In recent experiments a new case was identified in which the weak interactions of an acetyl group with a gold adatom serves to direct dispersion forces for bringing acetylbiphenyl (ABP) molecules together and initiate the on-surface assembly of acetylbiphenyl supramolecular structures\cite{nickel_moving_2013,ohmann_supramolecular_2015,
eisenhut_training_2016} Up to four ABP molecules can be weakly bound in this way. Three ingredients are needed to create these supramolecular structures. First, weak long-range dispersion forces are necessary to bring the ABP molecules together. Second, a passive supporting surface is important for the ABP molecules to easily diffuse at room temperature to nucleate and form the supramolecules. And third, metal adatoms must be present and diffuse on the surface to be captured by one ABP molecule facilitating the attachment of the other molecules to form a supramolecule.

\section{Methods}
\subsection{Theory}
Calculations have been performed in the framework of density functional theory as implemented in the VASP code\cite{kresse_efficiency_1996} using the projected augmented-wave (PAW) method.\cite{kresse_ultrasoft_1999} Wave functions have been expanded using a plane wave basis set with an energy cut-off of 400 eV. We used  PBE as GGA-funcional.\cite{perdew_generalized_1996} This functional was completed with dispersion corrections introduced through the Tkatchenko-Scheffler scheme.\cite{tkatchenko_accurate_2009} The Au (111) surface was modeled using a slab geometry with a surface unit cell $12\times 6\sqrt{3}$ with four Au layers and a vacuum region of 13~\AA. The active calculation system contains 684 atoms, 576 of them were gold atoms modeled with an 11-active-electron PAW. These heavy calculations require massive computational resources. We have kept the bottom two Au layers fixed, allowing the two topmost layers and the supported molecules to relax until forces were smaller than 0.02 eV/\AA. Despite the large supercell, a $3\times3\times1$ k-mesh had to be used in order to get converged results for the dI/dV maps. Charge transfer has been obtained by performing a Bader analysis.\cite{tang_grid-based_2009} STM topographic images have been simulated applying the Tersoff and Hamann approximation,\cite{tersoff_theory_1983,tersoff_theory_1985} using the method described by Bocquet et al.\cite{bocquet_theory_2009} as implemented in STMpw.\cite{lorente_stmpw_2019}
The binding energies have been calculated as
\begin{multline}
E_{bind}=4\cdot E(ABP@Au111) + E(m\cdot Au_{ad}+Au111) -\\ 4\cdot E(Au111)- E(4\cdot ABP@(m\cdot Au_{ad}+Au111)),
\end{multline}
where $E(ABP@Au111)$ is the energy of a supported ABP molecule; $E(m\cdot Au_{ad}+Au111)$ is the energy of the (111) surface of Au with $m$ Au adatoms; $E(Au111)$ is the energy of the (111) surface of Au; and $E(4\cdot ABP@(m\cdot Au_{ad}+Au111))$ is the energy of the supported tetramer including $m$ Au adatoms. 
The adsorption energies have been calculated as
\begin{multline} 
E_{ad}=[n\cdot E(ABP) + E(Au111)-\\ E((n\cdot ABP)@(m\cdot Au_{ad} + Au111))]/n,
\end{multline} 
where $n$ is the number of ABP molecules and $E(ABP)$ is energy for the gas phase ABP molecule. 
The adsorption Gibbs free energy per area unit, $\Delta g_a(\Delta\mu(T,P))$ has been calculated as a function of the gas-phase chemical potential, $\Delta\mu(T,P)$, for a given gas temperature T, and pressure P.\cite{reuter_composition_2001} For $n$ molecules on a unit area A, the adsorption free energy can be approximated by:
\begin{equation}
\Delta g_a(\Delta\mu(T,P))\approx n/A (\varepsilon-\Delta\mu(T,P)) 
\end{equation}
This expression neglects zero point energies, and vibrational differences between the adsorbed and the free molecules, because these differences are negligible in front of the adsorption interactions. The mean adsorption energy per adatom is calculated as
\begin{multline} 
\varepsilon = ( E((tetramer+n\cdot Au)@Au(111)) -\\ E(tetramer@Au(111))- n\cdot E(Au_{gas}) )/n 
\end{multline}

\subsection{Experiment}

4-acetylbiphenyl (ABP) molecules were deposited at 320 K from a Knudsen cell onto the Au(111) substrate kept at room temperature.\cite{nickel_moving_2013} The sample was cooled down to 5 K after this deposition and low temperature STM experiments were performed. The STM images show a rich variety of molecular structures formed during the sublimation. Single ABP, double, triple and quadruple ABP supramolecules can be identified on the Au(111) surface terraces.\cite{eisenhut_training_2016} By slightly annealing the surface after deposition to about 50$^{\circ}$C, the quadruple structure becomes dominant.\cite{ohmann_supramolecular_2015}\cite{eisenhut_training_2016} 

\section{Results and discussion}
\begin{figure*}
 \centering
 \includegraphics[height=9cm]{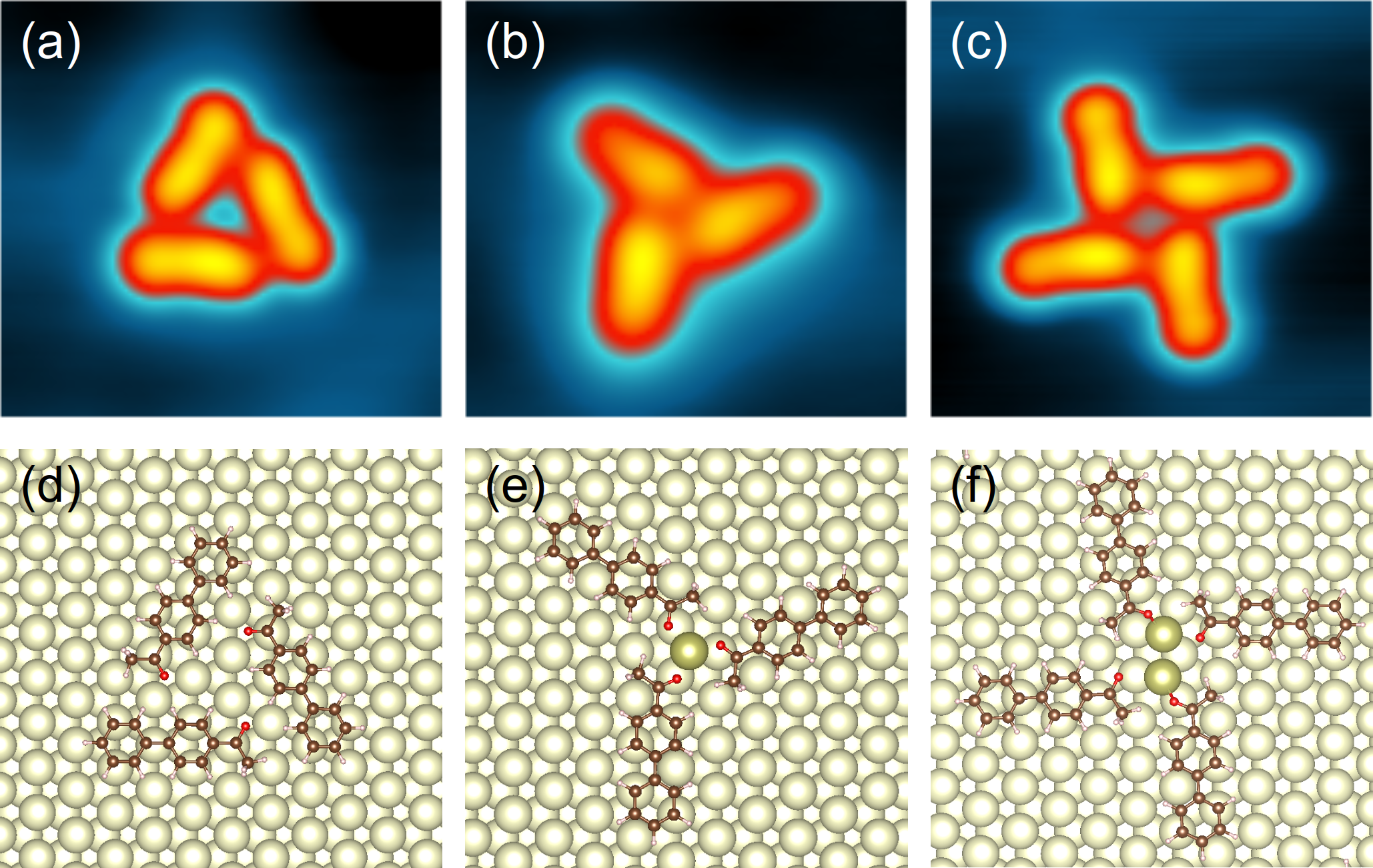}
 \caption{Experimental constant current STM images of the observed ABP supramolecular structures: (a) trimer, (b) intersecting trimer, (c) intersecting tetramer. Images size: 3.9 nm x 3.9 nm, V = 100 mV, I = 50 pA. (d-f) Corresponding ball-and-stick molecular structures optimized by DFT. Note that the proposed structures (e) and (f) include adatoms (darker golden balls).}
 \label{fig1}
\end{figure*}

Fig.~\ref{fig1} presents the experimental STM images of the most common supramolecules:\cite{eisenhut_training_2016} a trimer, an intersecting trimer and an intersecting tetramer, and the corresponding ball-and-stick molecular structures optimized by density functional theory (DFT).

Single ABP molecules have been studied in detail in a previous work.\cite{zobac_directionality_2020} ABP molecules are physisorbed by van der Waals forces on Au(111) with apparently no need to be stabilized by a single gold adatom. Extensive calculations using DFT confirmed that there is no charge transfer between a single ABP and the Au(111) surface in this case. There is however a very small polarization of the substrate electronic cloud due to the presence of a small ABP molecular dipole.\cite{zobac_directionality_2020} 

The first ABP trimer configuration (Fig.~\ref{fig1}a) presents a triangular supramolecular structure indicating a maximization of the H-bonding interactions between the edge O-atom of one ABP and the phenyl H atoms on the next one. Albeit weak, those interactions are large enough to stabilize the diffusion barrier of ABP trimers as compared to single ABP molecules on Au(111) without the need of gold adatoms. DFT calculations using semi-empirical van-der-Waals corrections show that the largest contribution to the interaction energy for the trimer stabilization comes from the van-der-Waals interactions between the ABP molecules. The final ABP orientation in the trimer is given by the very weak O-H-C bond constrained by the Au(111) surface crystallographic directions. The acetyl groups alternate along this triangular structure leading to the on-surface induced ABP trimer chirality. Notice that the same molecular structure has been observed with OH-terminated biphenyl molecules.\cite{kong_long-range_2020} This structure is different from the one found in the tetramer and leads to H-bonds that are stronger than in the tetramer where only methyl-oxygen bonds seem to be operative. Although generally the H-bond is much weaker for ABP molecules, it is strong enough to direct the formation of the trimer and gives it a definite chirality as compared to the ABP Au(111) surface lateral energy diffusion barrier.

The second ABP trimer supramolecular structure (Fig.~\ref{fig1}b) was not found for OH-substituted biphenyl molecular trimer structures. It is a star like intersecting ABP structure in such a way that the ABP molecules minimize their O-O mutual distances also leading to a chiral structure. In such a structure, the CH3-O hydrogen bond in between two ABP molecules is extremely weak and definitely weaker than the OH case. It is usually recognized as not existing, although its strength depends on the molecular environment.\cite{yesselman_frequent_2015} This new trimer structure is also stabilized by long-range dispersive forces. However, its actual shape cannot be explained by such hydrogen bond alone and at least the capture of a surface gold adatom is required (see discussion). Although the triangular ABP trimer (Fig.~\ref{fig1}a) cannot be altered by STM tip pulses as large as 2 V, the intersecting ABP trimer (Fig.~\ref{fig1}b) can be manipulated by voltage pulses as an independent supramolecular structure.\cite{eisenhut_training_2016}

Unique to ABP molecules is their possibility to form tetramers on Au(111) stabilized via their acetyl group, as presented in Fig.~\ref{fig1}c. In comparison to the trimer case (Fig.~\ref{fig1}b) the contribution of the hydrogen bonds to the tetramer stability is extremely small. Notice that the windmill-shaped tetramers are reproducibly observed over the Au(111) surface and can reach large densities, approaching the monolayer (Electronic Supplementary Information (ESI) Fig. S1). The tetramers also have a clear on-surface induced chirality. They are cohesive enough to be moved on the surface using 2 V bias voltage pulses [9],\cite{nickel_moving_2013} which is beneficial for single molecule mechanics studies [9-11].\cite{nickel_moving_2013,ohmann_supramolecular_2015,eisenhut_training_2016}

The van der Waals interactions between single ABP and the Au(111) surface sets a clear directionality for ABP physisorption.\cite{zobac_directionality_2020} ABP preferentially adsorbs along the [$1\bar{1}0$] and [$11\bar{2}$] directions of the Au(111) surface to maximize the interaction of the phenyl rings on hollow sites.\cite{liu_benzene_2012} This is confirmed by the STM images of the different supramolecular structures presented in Fig.~\ref{fig1} where the ABP molecules are uniquely aligned along these two directions.

At room temperature, gold adatoms are known to diffuse on the Au(111) surface, detaching for example from step edges at room temperature.\cite{jaklevic_scanning-tunneling-microscope_1988} They have been observed to bond to different molecules adsorbed on Au(111)\cite{maksymovych_gold-adatom-mediated_2006,gronbeck_goldthiolate_2008,shi_porphyrin-based_2009,boscoboinik_one-dimensional_2010,yang_orbital_2014,
maksymovych_gold_2010} where organic molecules easily capture them.\cite{meyer_tuning_2015,maksymovych_gold-adatom-mediated_2006,
gronbeck_goldthiolate_2008,shi_porphyrin-based_2009,boscoboinik_one-dimensional_2010,yang_orbital_2014,maksymovych_gold_2010} In some instances, isolated gold adatoms serve as the nucleation centers for the self-assembly of molecular nanostructures.\cite{meyer_tuning_2015} 
\begin{figure}
 \centering
 \includegraphics[width=\columnwidth]{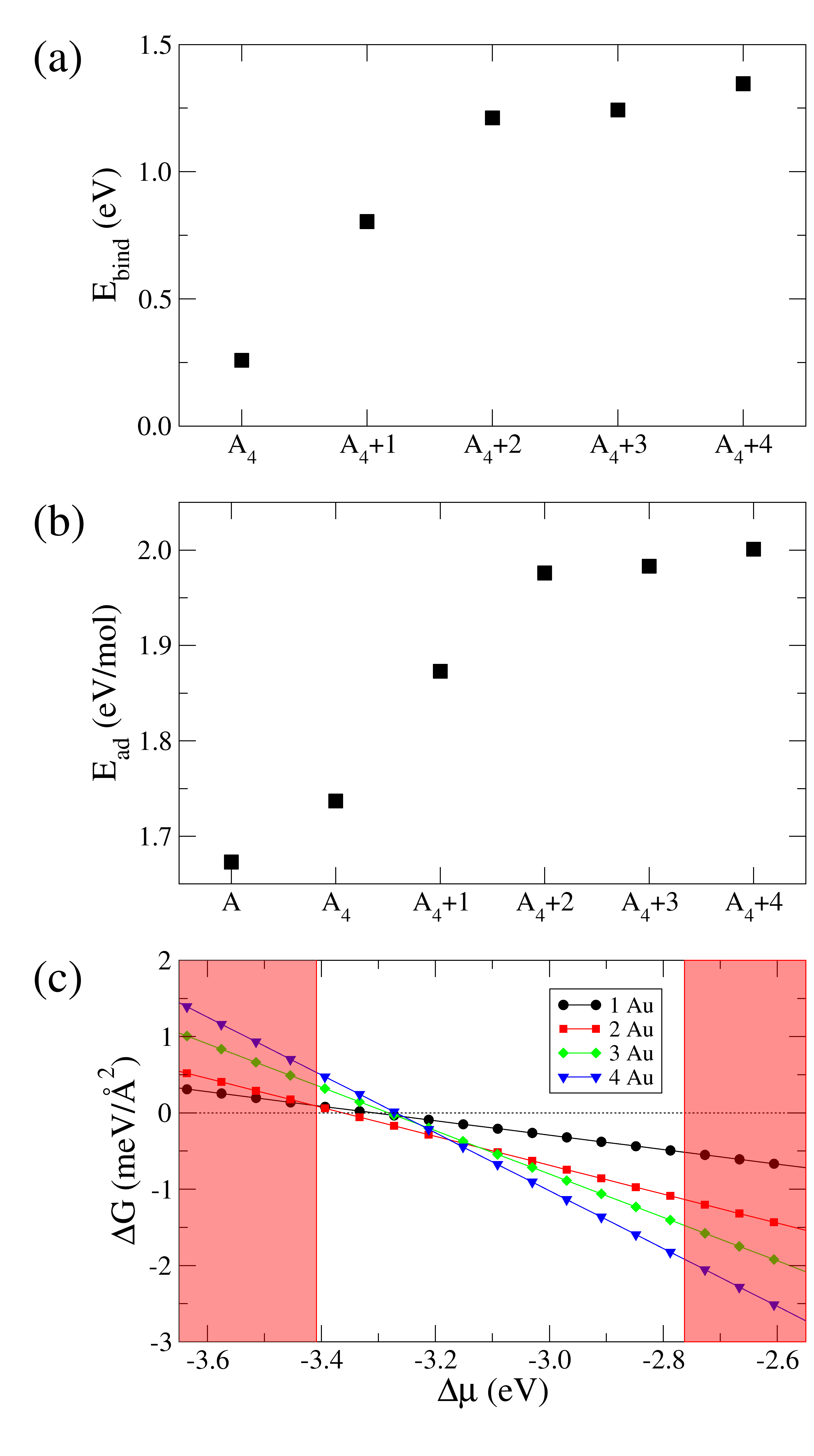}
 \caption{The energetics of supramolecular formation. (a) Binding energy of ABP molecules in a 4-molecule supramolecular structure as a function of the number of adatoms in the supramolecule. The ABP molecule is denoted by A and the tetramer plus $m$ adatoms is A4+$m$ (b) Adsorption energy per ABP molecule of the 4-molecule system as a function of adatoms. (c) Gibbs free energy as a function of the chemical potential of a source of adatoms. The red areas are considered not to be accessible because they correspond to chemical potentials of adatoms that are either smaller than the chemical potential of a bulk source of adatoms (left line) or bigger than the surface source (right line) that are limiting cases for possible sources of adatoms.}
 \label{fig2}
\end{figure}
Using DFT we have calculated how a single ABP molecule binds to a gold adatom on the Au(111) surface. We obtained that the binding energy of ABP increases by 377 meV as compared to the case without a gold adatom. Such ABP-Au complex further randomly diffuses on the Au(111) surface at room temperature and can work as a nucleation center for supramolecular structures. To confirm this point, in the absence of surface and without a central gold atom, we have first calculated the binding energy of 4 ABP molecules in the gas phase. Their van der Waals interactions lead to a total binding energy of ~300 meV. It is worth noting that the substitution of their acetyl groups by hydroxyl groups leads to a binding energy more than five times larger, confirming the strength of an O-H-O hydrogen bond. When now the tetramer is adsorbed on Au(111) still without gold adatoms, the binding energy gain of the 4 ABP molecules with respect to 4 ABP molecules considered separately adsorbed on the Au(111) surface is 259 meV. Despite the difference in geometry between the free of surface 2D conformation and the surface conformation, this value is very close to the gas phase one because of the binding through van der Waals forces. However, in the presence of a central gold adatom, the binding energy increases up to 804 meV as presented in Fig.~\ref{fig2}a. Notice also that, as presented in Fig.~\ref{fig2}b, the adsorption energy per ABP molecule increases from 1.74 eV/molecule to 1.87 eV/molecule in the presence of one gold atom in the center of the supramolecule.

It is important to note that a gold adatom does not form a covalent bond with the ABP molecules. It rather induces a redistribution of charges among the ABP molecules leading to the supramolecule structure stabilization. This can be observed on the LDOS where the molecular resonances are shifted to lower energies and broaden but without charge transfer (ESI Fig. S2b). A Bader charge analysis yields that one single ABP in the supramolecule transfers 0.075 electrons to the substrate. With two gold adatoms, 0.092 electrons per molecule are transferred.

In order to investigate the plausibility of the involvement of gold adatoms in the formation of the ABP supramolecules we computed the changes in Gibbs free energy as a function of the chemical potential of the source of gold atoms. If the change in free energy is positive, the gold adatoms will stay at the source. However, if it is negative the adatoms will become part of the ABP supramolecules. Fig.~\ref{fig2}c shows the variation of the Gibbs energy as a function of the atom chemical potential. In order to have realistic references, we have assumed two limits for the source of Au adatoms. In the first case, the gold atoms are strongly bound to bulk gold. The corresponding chemical potential is taken as the energy of a bulk atom (left vertical line in Fig.~\ref{fig2}c). The other extreme is the binding energy of a gold adatom to the gold surface (right vertical line), given that gold adatoms diffuse on the Au(111) surface at room temperature.\cite{jaklevic_scanning-tunneling-microscope_1988} The chemical potentials of these or other sources of Au adatoms (herringbone reconstructions, step edges...) are therefore most likely lying between these two limits. Fig.~\ref{fig2}c shows that, within the two limits, phases with zero, two or four adatoms in the supramolecules can exist as the chemical potential increases.
\begin{figure*}
 \centering
 \includegraphics[height=9cm]{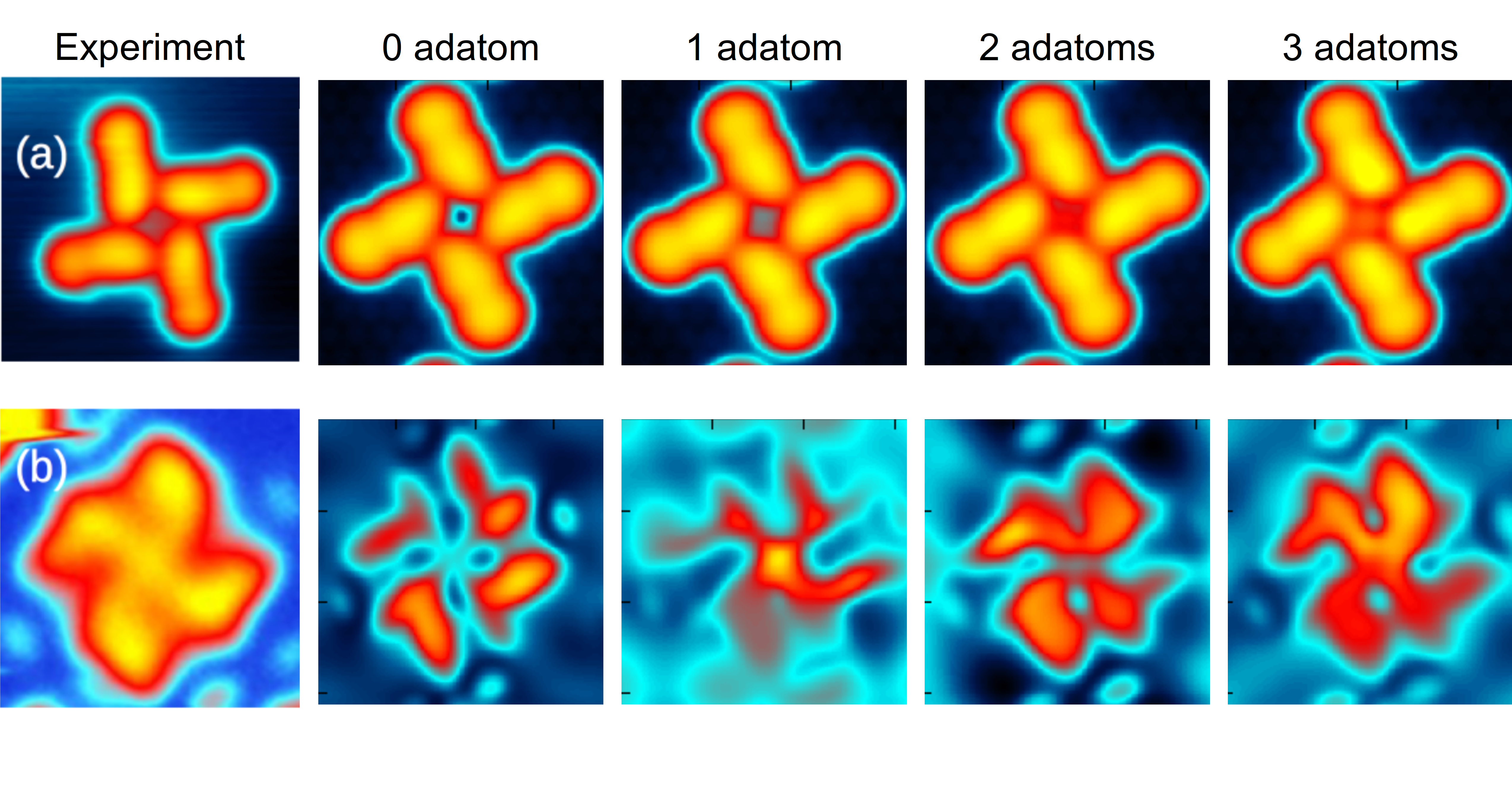}
 \caption{Upper row (a) Comparison experiment-simulation of STM images; a typical experimental STM image of an ABP tetramer and four simulated images with respectively  0, 1, 2 and 3 adatoms in the center of the ABP tetramer structure. Lower row (b) a typical experimental dI/dV map of an ABP tetramer and the corresponding simulated dI/dV maps. The experimental parameters were, for the dI/dV map at 2.15 V, I = 25 pA, for the topographic image. V$_{sample}$= -1 V, I = 190 pA. The theoretical biases to obtain the dI/dV maps are 1.62, 1.40, 1.27 and 1.21 V, respectively.}
 \label{fig3}
\end{figure*}
Further structural information about the conformation of ABP tetramers on the surface can be obtained from a comparison between experimental and simulated constant current STM images. Fig.~\ref{fig3}a shows the simulated STM images for tetramers with 0 to 3 gold adatoms positioned at the center of the structure and the corresponding  STM experimental image. By comparing the contrast of the STM images, the best matching is obtained with one single gold adatom. Indeed, defining in the STM image z0 as the depth of the contrast depression at the center of the ABP molecular structure and L as the maximum corrugation over a tetramer, experimental data consistently give z0/L between 45 and 50\%. The z0/L coming from STM image simulation gives 57\% for zero, 44\% for one, 35\% for two and 30\% for three gold adatoms. Let us also note that a gold adatom in a different position from the center produces a clear signature in the simulated STM image (ESI Fig.~S3), in agreement with experimental observations [10]. Differential conductance maps can also be used to perform such a comparison. In this case, as presented in Fig.~\ref{fig3}b, the best agreement is reached for the two or three gold adatoms cases. From all these comparisons, the presence of at least one gold adatom at the center of the ABP tetramer seems likely for the tetramers (Fig.~\ref{fig1}c)  and for the intersecting trimer (Fig.~\ref{fig1}b, see ESI Fig.~S4). 

Previous work had shown that a supramolecular tetramer can be manipulated using STM tip bias pulses.\cite{nickel_moving_2013,ohmann_supramolecular_2015,eisenhut_training_2016} The presence of adatoms increases the electronic density of states of the tetramer at empty and occupied states thus increasing the tunneling current intensity through this supramolecule.  Moreover, the polarity dependence of the manipulations\cite{nickel_moving_2013} hints at electrostatic interactions between tip apex and the tetramer supramolecule. This seems unlikely in the absence of adatoms given the very small charge transfer with the substrate. However, nudged-elastic-band calculations yield diffusion barriers that are consistently higher for supramolecules containing adatoms. Indeed the barrier for diffusion along the [$11\bar{2}$] direction of a supramolecule without adatoms is 243 meV, while with two adatoms the barrier becomes 347 meV [ESI Fig.~S5 and S6]. Yet, the manipulations are performed using pulses around 2 V. At this voltage, the electrons are tunneling through the tetramer with energies well above the diffusion barriers. It is interesting to realize that only when adatoms are present, the supramolecular binding energy is noticeably larger than the diffusion barrier.

\section{Conclusions}
\balance
Supramolecular chemistry is based on weak interactions among molecular building blocks. We have analyzed here how this can be performed on a supporting surface. The assembly of ABP molecules on Au(111) is an interesting example of a supramolecule of limited lateral size stabilized by very weak interactions that responds without any decomposition to external perturbations like STM voltage pulses. Those interactions are originated from long-range dispersive forces between the ABP and the Au(111) surface respecting also surface crystalline orientation for the assembly of the dimer, trimers and tetramers. The stabilization agents for the peculiar windmill- like ABP tetramers are the native gold adatoms diffusing at room temperature randomly on the Au(111) surface. They are reactive enough to serve as nucleation centers for the assembly of ABP trimers and tetramers, respecting also the Au(111) crystalline surface orientation. Our analysis suggests new ways of creating complex but small nano-objects that can eventually be used as devices or as seeds for extended hierarchical structures.

\section*{Conflicts of interest}
There are no conflicts to declare.

\section*{Acknowledgements}
We are grateful for funding from the EU-FET Open H2020 Mechanics with Molecules project (grant 766864). The authors thankfully acknowledge the computer resources at Finisterrae II (RES-QCM-2019-1-0024) and Cibeles (RES-QS-2019-3-0012) and the technical support provided by CESGA and CCC-UAM. 




\bibliography{references} 
\bibliographystyle{rsc} 

\end{document}